\documentstyle[aps,12pt]{revtex}

\begin{document}
\author{J.P. Krisch and E.N. Glass\thanks{%
Permanent address: Physics Department, University of Windsor, Ontario N9B
3P4, Canada}}
\address{Department of Physics, University of Michigan, Ann Arbor, Michgan 48109}
\date{31 January 2001, \ \ \ preprint MCTP-01-15}
\title{Dimension in a Radiative Stellar Atmosphere\thanks{%
The original version of this paper received Honorable Mention as a 2000
Gravity Research Foundation Essay} }
\maketitle

\begin{abstract}
------------------------------------------------------------------------------------------

Dimensional scales are examined in an extended 3+1 Vaidya atmosphere
surrounding a Schwarzschild source. At one scale, the Vaidya null fluid
vanishes and the spacetime contains only a single spherical 2-surface. Both
of these behaviors can be addressed by including higher dimensions in the
spacetime metric.

-------------------------------------------------------------------------------------------

\bigskip

\newpage\ \newline
\end{abstract}

\section{INTRODUCTION}

The nature and number of dimensions of the Universe are evolving ideas. From
the $3+1$ description of Minkowski spacetime, we have come to study theories
involving much higher dimensions \cite{Abel99},\cite{Duf99},\cite{KS98}, 
\cite{GZ99},\cite{LW98}. There are descriptions of processes requiring
non-integer dimensions \cite{Man82} and negative dimensions can give
interesting results in some calculations \cite{Dup88}. Spacetime models in
lower dimensions have proved useful in gaining insights to some of the
problems of quantum gravity \cite{Car95}. Identifying observable dimensional
effects is one of the critical components to understanding our universe.

The Vaidya spacetime provides a useful laboratory for studying possible
effects of fractal dimensions. A generalized Vaidya spacetime with horizon
function $m_{(d)}(u)$ and dimension parameter $d$ has metric \cite{GK00b} 
\begin{equation}
ds_{3+1}^{2}=g_{ab}^{Vad}(d)\,dx^{a}dx^{b}=[1-\frac{2m_{(d)}}{r}%
]du^{2}+2dudr-(r/r_{0})^{d-1}r_{0}^{2}d\Omega ^{2},  \label{d-met}
\end{equation}
where $r_{0}$ has units of length so that spheres have area of (length)$^{2}$%
. As we examine radial lines at fractal scales, each $r$ value provides a
set of points enclosed by a 2-sphere with luminosity area\ proportional to $%
r^{d-1}$. The metric describes a Schwarzschild object surrounded by a two
fluid atmosphere; a radiation fluid and a fluid with density $\Gamma ,$
radial heat flow $q_{r}$, and radial and transverse stresses \cite{GK00b}, 
\cite{GK99}. The atmosphere supports a variety of processes occurring at
microscopic scales, and for each of these processes, the metric can be
written with an appropriate dimension, $d$. For example, with $d=3$ and $%
m_{(3)}(u)=m_{0}$, an observer sees the Schwarzschild vacuum. The same
observer, looking at smaller scales, i.e. at different $d$ in the same
physical spacetime, would see processes characterized by a fractal
dimension, different $m_{(d)}$ and non-zero fluid content. If we are
observing in a $3+1$ spacetime, the horizon function $m_{(d)}(u)$ measures
the system mass only for spatial dimension $d=3$. For all $d$, the sectional
curvature mass $M$ \cite{GK99},\cite{GK00a} and density $\Gamma $ describe
the content within 2-surfaces of constant $u$ and $r:$ 
\begin{eqnarray}
-\frac{2M}{r^{3}} &=&R_{abcd}\hat{\vartheta}^{a}\hat{\varphi}^{b}\hat{%
\vartheta}^{c}\hat{\varphi}^{d}=\frac{(d-1)^{2}}{4r^{2}}[1-\frac{2m_{(d)}}{r}%
]-\frac{r_{0}^{d-3}}{r^{d-1}},  \label{sec-m} \\
4\pi (\Gamma -\Gamma _{0}) &=&r_{0}^{d-3}r^{1-d}\partial _{r}M
\label{sec-dens}
\end{eqnarray}
where $\hat{\vartheta}^{a}=r_{0}^{(d-3)/2}r^{(1-d)/2}\ \delta _{\vartheta
}^{a}$ and $\hat{\varphi}^{a}=r_{0}^{(d-3)/2}r^{(1-d)/2}\sin ^{-1}\vartheta
\ \delta _{\varphi }^{a}$.

The Raychaudhuri equation, used with null outgoing or incoming generators,
determines the focusing of null rays. The existence of an horizon is
expected to persist at all scales, since the focusing properties of incoming
and outgoing null geodesics transform covariantly with the size of their
orthogonal 2-spheres.

The interpretation of the metric parameter $d$ as a fractal dimension
follows from considering diffusive processes in the atmosphere. If ${\cal P}%
(u,r)$ is a quantity that is diffusing, the diffusion equation is 
\begin{equation}
\partial _{u}{\cal P}=\frac{1}{r^{d-1}}\partial _{r}(Dr^{d-1}\partial _{r}%
{\cal P})  \label{prob}
\end{equation}
and $d$ is identified as the dimension of the fractal substrate on which the
diffusion is occurring \cite{OP85},\cite{DMS94}. At smaller quantum levels, $%
{\cal P}$ is a probability measure on the space of radial paths \cite{Nel85}%
. Fractal dimensions are defined and discussed by Hughes \cite{Hug95}.
Diffusion is a probe of the smaller fractal dimensions because a continuous
description of a diffusive process can envelope underlying dimensional
behavior \cite{OP85}$.$ Diffusion observations can provide information about
these smaller scales which are not directly accessible to observation.

\section{DIMENSION AND\ FLUID\ CONTENT}

The field equations provide a 2-fluid description of the matter and
radiation content \cite{GK99} 
\begin{mathletters}
\begin{eqnarray}
8\pi \rho _{null} &=&-\frac{(d-1)\partial _{u}m_{(d)}}{r^{2}}+\frac{%
(d-1)(3-d)}{8r^{2}}[1-\frac{2m_{(d)}}{r}],  \label{rho-n} \\
8\pi \rho &=&\frac{r_{0}^{d-3}}{r^{d-1}}+\frac{(d-1)}{8r^{2}}[11-5d+\frac{%
5(d-3)m_{(d)}}{r}],  \label{rho} \\
8\pi (\rho +p_{r}) &=&16\pi q_{r}=\frac{(3-d)(d-1)}{4r^{2}}[1-\frac{2m_{(d)}%
}{r}],  \label{rho-pr} \\
8\pi p_{\bot } &=&-\frac{(d-3)}{r^{2}}[\frac{(d-3)m_{(d)}}{2r}+\frac{d-1}{%
4r^{2}}].  \label{p-perp}
\end{eqnarray}
$\rho $ can be related to $\Gamma $ through $M$. For dimension $d=3$, we
have the equivalences $M=m_{(3)}$ and $\Gamma =\rho $.

There are some interesting aspects to the $d=1$ scale. The sectional
curvature mass and density are $M=M_{0}r^{3},$ and $\Gamma =\Gamma _{0}r^{2}$%
. The fluid content is 
\end{mathletters}
\begin{mathletters}
\begin{eqnarray}
\rho _{null} &=&q_{r}=0, \\
\rho &=&-p_{r}=\frac{1}{8\pi r_{0}^{2}}, \\
8\pi p_{\bot } &=&-\frac{2m_{(1)}}{r^{3}}.
\end{eqnarray}
The null radiation and heat flow vanish while the density and radial stress
assume constant values. From the metric 
\end{mathletters}
\begin{equation}
ds_{3+1}^{2}=[1-\frac{2m_{(1)}}{r}]du^{2}+2dudr-r_{0}^{2}d\Omega ^{2}
\label{d1-met}
\end{equation}
one sees that all spheres have the same constant surface area $4\pi
r_{0}^{2} $. In the range of spacetime scales, $d=1$ marks the boundary
between a geometry where surface area and circumference increase with radius
and one where they decrease with radius.

\section{INVARIANTS}

As an aid to identifying the coordinate independent behavior of fractal
dimension $d$, we compute tensor invariants of metric (\ref{d-met}). The
Ricci scalar is given by 
\begin{equation}
R_{ab}g^{ab}=\frac{2}{r^{2}}\left( \frac{r_{0}}{r}\right) ^{d-3}+\frac{%
m_{(d)}(d-3)(3d-5)}{r^{3}}+\frac{(1-d)(3d-7)}{2r^{2}}.  \label{ricci-sc}
\end{equation}
The Ricci, Weyl, and Riemann quadratic invariants are 
\begin{eqnarray}
R_{ab}R^{ab} &=&\frac{2}{r^{4}}+(\frac{2}{r_{0}^{4}})\left( \frac{r_{0}}{r}%
\right) ^{2d-2}+2(\frac{d-1}{r_{0}^{5}})\left( \frac{r_{0}}{r}\right) ^{d+2}%
\left[ 2m_{(d)}(d-3)-r(d-2)\right]  \label{riccisq} \\
&&+(\frac{d-3}{4r^{6}})\left[ 
\begin{array}{c}
4m_{(d)}^{2}(d-3)(3d^{2}-8d+7)+4\partial _{u}m_{(d)}r^{2}(d-1)^{2} \\ 
+r^{2}d(3d^{2}-11d+15)-dm_{(d)}r(12d^{2}-56d+84)+r(40m_{(d)}-3r)
\end{array}
\right] .  \nonumber
\end{eqnarray}
\begin{equation}
C_{abcd}C^{abcd}=\frac{1}{3r^{4}}\left[ d-1-4dm_{(d)}/r-2(r_{0}/r)^{d-3}%
\right] ^{2}.  \label{weylsq}
\end{equation}
\begin{eqnarray}
R_{abcd}R^{abcd} &=&\left( \frac{2}{r_{0}^{2}}\right) ^{2}\left( \frac{r_{0}%
}{r}\right) ^{2d-2}+\left( \frac{m_{(d)}}{r^{3}}\right)
^{2}[(d-3)(3d^{3}-15d^{2}+29d-1)+48]  \label{riemsq} \\
&&+\left( \frac{d-1}{2r^{2}}\right) ^{2}(3d^{2}-14d+19)+\left( \frac{%
2\partial _{u}m_{(d)}}{r^{4}}\right) (d-3)(d-1)^{2}  \nonumber \\
&&-\left( \frac{m_{(d)}}{r^{5}}\right) (d-1)^{2}(3d^{2}-16d+25)+\left[ \frac{%
2(d-1)^{2}}{r^{5}}\right] \left( \frac{r_{0}}{r}\right) ^{d-3}(2m_{(d)}-r). 
\nonumber
\end{eqnarray}
Values of the invariant scalars are tabulated for the $d=3$ Vaidya null
fluid and the $d=1$ single sphere space ($S_{ab}=R_{ab}-%
{\frac14}%
Rg_{ab}$ is the trace-free Ricci tensor):

$
\begin{array}{ccc}
& \underline{d=1} & \underline{d=3} \\ 
\underline{R_{ab}g^{ab}\ \ \ \ \ } & \ \ \ \ \ 2/r_{0}^{2}+4m_{(1)}/r^{3} & 
\ \ \ \ \ \ 0 \\ 
\underline{R_{ab}R^{ab}\ \ \ \ } & \ \ \ \ \ \ \ \
2/r_{0}^{4}+8m_{(1)}^{2}/r^{6}\ \ \ \  & \ \ \ \ \ \ 0 \\ 
\underline{C_{abcd}C^{abcd}} & \ \ \ \ \ \
(1/3)[2/r_{0}^{2}+4m_{(1)}/r^{3}]^{2} & \ \ \ \ \ \ 48m_{(3)}^{2}/r^{6} \\ 
\underline{R_{abcd}R^{abcd}} & \ \ \ \ \ \ \ \
4/r_{0}^{4}+16m_{(1)}^{2}/r^{6}\ \ \ \  & \ \ \ \ \ \ 48m_{(3)}^{2}/r^{6} \\ 
\underline{S_{ab}S^{ab}\ \ \ \ } & \ \ \ \ \ \ \
[1/r_{0}^{2}-2m_{(1)}/r^{3}]^{2}\ \ \ \  & \ \ \ \ \ \ 0 \\ 
\underline{S_{ab}S_{\ c}^{b}S^{ac}} & \ \ \ \ \ \ 0\ \ \ \ \  & \ \ \ \ \ \ 0
\end{array}
$\newline
\ \newline
A set of 16 real valued scalar invariants is given by Carminati and
McLenaghan \cite{CM91}. Using GRTensor \cite{MPL00} we have computed the
entire set for $d=1$ and $d=3$. The $d=3$ Vaidya spacetime has only 2
non-zero scalars out of 16, and they are powers of $m_{(3)}/r^{3}$, so $d=3$
has one independent invariant. The $d=1$ space has 9 non-zero
Carminati-McLenaghan scalars, all powers of $(r^{3}\pm
2r_{0}^{2}m_{(1)})/(r_{0}^{2}r^{3})$. Hence $d=1$ has two independent
invariants. None of the scalar invariants place the $d=1$ family in a
special category. However, when the optical scalars for the null generators
of 2-spheres are examined then $d=1$ stands out.

\section{OPTICAL SCALARS}

We write metric (\ref{d-met}) as $g_{ab}^{Vad}(d)=2l_{(a}n_{b)}-2m_{(a}\bar{m%
}_{b)}$ using a Newman-Penrose null tetrad where 
\begin{mathletters}
\begin{eqnarray}
l_{a}dx^{a} &=&du,  \label{tet_a} \\
n_{a}dx^{a} &=&dr+(1/2)[1-\frac{2m_{(d)}}{r}]\ du,  \label{tet_b} \\
m_{a}dx^{a} &=&-(r_{0}/\sqrt{2})(r/r_{0})^{(d-1)/2}(d\vartheta +i\text{sin}%
\vartheta \ d\varphi ),  \label{tet_c}
\end{eqnarray}
with non-zero spin coefficients 
\end{mathletters}
\begin{mathletters}
\begin{eqnarray}
\rho &=&(1-d)/(2r),\ \ \ \ \mu =[(1-d)/(4r)][1-\frac{2m_{(d)}}{r}],
\label{tet-a} \\
\alpha &=&-(r_{0}/r)^{(d-1)/2}\text{cot}\vartheta /(2\sqrt{2}r_{0})=-\beta ,
\label{tet-b} \\
\gamma &=&m_{(d)}/(2r^{2}).  \label{tet-c}
\end{eqnarray}
The only non-zero Weyl tensor component is 
\end{mathletters}
\begin{equation}
\Psi _{2}=-(\frac{1}{6r_{0}^{2}})\left( \frac{r_{0}}{r}\right) ^{d-1}+\frac{%
(d-1)r-4m_{(d)}d}{12r^{3}}  \label{psi2}
\end{equation}
identifying the $g_{ab}^{Vad}(d)$ spacetime as Petrov type {\bf D}. Both $%
l^{a}$ and $n^{a}$ are tangent to hypersurface orthogonal null geodesics.
Spin coefficients $\rho $ and $\mu $ are the optical scalars describing the
expansion of the incoming and outgoing null geodesic congruences from
two-surfaces of constant $u$ and $r$. When $d=1$, we see from Eq.(\ref{tet-a}%
) that the expansion is zero, and so the ($\vartheta ,\varphi $) two-surface
at $r_{0}$ is marginally trapped.

\section{POSSIBILITIES}

There are several possible explanations for the $d=1$ behavior. The first is
that $d$ must be greater than one. Eliminating $d\leq 1$ would be necessary
if the field equation description were not correct for this range of $d$.
The model describes $d$ as a fractal dimension which changes as one goes
down in spatial scales. At some scale one might expect fractal functions to
provide valid descriptions of physical parameters like a matter
distributions. Since fractal functions like the Weierstrass function \cite
{BL80} are continuous, but not differentiable, conventional general
relativity would not work \cite{Kob00}, \cite{ElN00}. Another possibility is
that $d=1$ is only observable in the final state. Lindquist, Schwartz, and
Misner \cite{LSM62} have pointed out that Vaidya's $u$ in metric (\ref{d-met}%
) covers the lower half-plane of Kruskal's ($v,w$) coordinates in Fig.(1) of 
\cite{LSM62}. The Schwarzschild ($T,r$) sector is the lower quadrant ($v\leq
0,$ $w\geq 0$). To quote from \cite{LSM62}: ''The hypersurface $%
r=2m_{(3)}(u\rightarrow \infty )$ in Vaidya's metric is analogous to the
Schwarzschild hypersurface $r=2m_{(3)}(T\rightarrow \infty )$ in Kruskal's
metric''.\ \ It is future null infinity that is the horizon at $r_{0}$. The
intersections of outgoing null surfaces with future null infinity are
2-spheres. Since there is only one 2-sphere at the $d=1$ scale, this implies
that $u\rightarrow \infty $ and the $d=1$ scale is only observable in the
late time Vaidya solution when the time dependence of $m_{(3)}(u)$ has
vanished. This is the second of the two possibilities mentioned by Lindquist
et al. We expect $m_{(3)}(u)$ will evolve toward $m_{(1)}(u\rightarrow
\infty )$. The existence of only a single spherical area suggests a new
possibility.

From the field equations we see that the density $\rho $ is constant for $%
d=1 $, and $\Gamma \sim r^{2}$. The behavior of these two densities can be
reconciled if $r=r_{0}$ and both $M$ and $\Gamma $ are constant. The
existence of only one spherical surface suggests that at the $d=1$ scale we
have lost a dimension and are looking only at the horizon. One of
Mandelbrot's examples\cite{Man82} of the variability of dimension describes
how a ball of thin thread is seen as an observer changes scale. From far
away it seems a point, which becomes a 3-dimensional ball at a closer
distance. As an observer moves down through various scales, the ball changes
to a set of 1-dimensional fibers which become 3-dimensional cylinders, etc.
While the embedding dimension for the ball has not changed, the effective
dimension of the contents does change. At $d=1$, the effective spatial
dimension of the Vaidya spacetime appears to be two. $d=1$ could mark a real
drop in dimension with the onset of new physics, or the apparent loss of
dimension may imply the existence of dimensions higher than four in the
''real'' spacetime. The new dimensions would be analogous to the embedding
dimensions in Mandelbrot's example. It is possible that there are compact 
\cite{YF99} or non-compact \cite{RS99} dimensions and one begins to see more
than four dimensions at $d=1$. At this scale, the $3+1$ metric is simply not
correct although general relativity can still provide a valid description.
Given the recent interest in higher dimensions in our universe this last
explanation is promising, requiring new dimensions but maintaining the
formalism of general relativity as an investigative tool.

\section{HIGHER DIMENSIONS}

Two features which emerge from the $3+1$ analysis are the vanishing of the
null fluid at the $d=1$ scale and the existence of a single 2-sphere for the
entire spacetime. Using conventional general relativity, both of these
features can be addressed with a higher dimensional metric. Labelling the
extra spatial dimension by coordinate $y$, one could have for example, 
\begin{eqnarray}
ds_{4+1}^{2} &=&[1-\frac{2m_{(d)}(u)}{r}%
]du^{2}+2dudr-(r/r_{0})^{d-1}r_{0}^{2}[d\vartheta ^{2}+\sin ^{2}\vartheta
d\varphi ^{2}]  \label{5d-met} \\
&&-(r/r_{1})^{2}dy^{2},  \nonumber
\end{eqnarray}
with $r_{1}$ a dimensional constant similar to $r_{0}$. The extra spatial
dimension allows higher dimensional spherical surfaces. Calculating the
field equations for $d=1$, the extra dimension provides a non-zero null
fluid density containing the term $-\partial _{u}m_{(1)}/r^{2}$. More
complicated metrics with $y$ dependant warp factors provide similar results.

The behavior of this system at $d=1$ could be evidence for higher dimensions
which begin to be important at the $d=1$ scale. Currently, some theories
involving small extra dimension have been studied; for example, the
Kaluza-Klein theories with a rolled up 5th dimension, superstring theories
with several methods of compactification \cite{GSW87}, the gravitational
bulk propagation model of Arkani-Hamed et al \cite{AHDD98} or the Cantorian
spacetime model described by El Naschie \cite{ElN98}. There have also been
various ways suggested to search for these extra dimensions.

If the scale of the extra dimensions is of order 1 mm, then there should be
evidence in torsion-balance tests \cite{LCP99}. Smaller scales might be
found in accelerator experiments. Scattering experiments which could find
evidence for extra dimensions have been discussed by groups at SLAC \cite
{GRW99} and CERN \cite{MPP99}. On the astrophysical level, Liu et al \cite
{LOC00} have discussed some solar system tests based on a 5-dimensional
extension of the Schwarzschild metric and Cassisi et al \cite{CCD+00} have
examined the effects of higher dimensions on stellar evolution. A lower
bound on extra dimensions based on light-cone fluctuations has been
discussed by Yu and Ford \cite{YF99}. Evidence for higher dimensions could
also show up in a transport process like diffusion.

Consider, for example, a diffusive process: let $p(u,r)$ be the probability
at time $u$ for an element of the atmosphere to be in a shell between $r$
and $r+dr$. $p(u,r)$ obeys an anomalous diffusion equation \cite{OP85} 
\begin{equation}
\partial _{u}p=\frac{1}{\sqrt{-g}}\partial _{r}[D\sqrt{-g}\partial _{r}p]
\label{diff-p}
\end{equation}
with $\sqrt{-g}\sim r^{d}$ and diffusivity $D(r)$\thinspace =$%
\,D_{0}r^{-\theta }.$ The solution is 
\begin{equation}
p(u,r)\sim u^{-(d+1)/(2+\theta )}e^{-cr^{2+\theta }/u}.  \label{p-soln}
\end{equation}
From this probability one can show that $<r^{2}(u)>$\ $\sim u^{2/(2+\theta
)} $ \cite{OP85}. The scaling with time samples the connectivity of the
substrate through the radial dependence of the diffusivity. There is
evidence, in condensed matter calculations, that the exponent $\theta $
depends on the number of spatial dimensions \cite{GAA83}. If this trend
carries over to gravitational problems, we would expect to see decreasing
time dependence in the scaling relation as the number of dimensions
increases. Observations of Vaidya atmospheric scaling behavior could provide
information about dimension. Focusing on physical quantities such as fluid
density, that are proportional to the probability density, would also allow
dimensional effects to be studied.

In conclusion, examining the $3+1$ Vaidya spacetime over decreasing scales
has highlighted a scale with features that can be explained, within
conventional relativity, by including higher dimensions. Dimensional physics
has become an active area of investigation with some promise of future
experimental insights \cite{Abel99}.

\end{document}